\documentclass[12pt]{article} 
\usepackage{amssymb}
\usepackage{graphicx}
\begin{document} 
\begin{flushright}
e-Print: arXiv:1010.0918 [hep-ph] \\
CERN-PH-TH/2010-221 \\
Pisma ZhETF  {\bf 92} (2010) 720 \\
$[$JETP Letters {\bf 92} (2010) 652$]$ 
\end{flushright}
\begin{center}
          {\large \bf Towards  a common origin \\ of the elliptic flow, ridge and alignment  } 

\vspace{0.5cm}                   
{\bf Igor M. Dremin$^1$  and  Victor T. Kim$^2$}

\vspace{0.5cm}              
         $^1$Lebedev Physical Institute, Moscow 119991, Russia\\

         $^2$St. Petersburg Nuclear Physics Institute, Gatchina 188300, Russia

\end{center}

\begin{abstract}
It is claimed that elliptic flow, ridge and alignment are effects of 
azimuthal asymmetry, which have a common origin evolving with primary energy
and stemming from the general structure of field-theoretical matrix elements.
It interrelates a new ridge-phenomenon, recently found at the LHC and RHIC, 
with known coplanarity feature observed in collider jet physics as well as  
in cosmic ray studies. 
\end{abstract}

Azimuthal asymmetry of multiparticle production in collisions of hadrons and 
nuclei was noticed in many experiments at various energies. It got the names
of elliptic flow, ridge and alignment (or coplanarity).

First it was measured in azimuthal distributions of nuclear fragments at rather
low energies of 2.5 GeV and 4.5 GeV per nucleon in Berkeley and Dubna (see, e.g.,
\cite{bon, nag}) when
studying nuclei collisions with emulsion nuclei. The positive values of the 
second Fourier coefficient of the series expansion of fragment distributions in differences
of the azimuthal angles have been obtained. Namely this parameter was later used
at the RHIC and called $v_2$ or elliptic flow:
\begin{equation}
E\frac {d^3N}{d^3p}=\frac {1}{2\pi }\frac {d^2N}{dy  \,  p_Tdp_T}
\Bigl( 1+ \sum _{n=1}^{\infty }2v_n\cos[n(\phi -\Psi _r)]  \Bigr), 
\label{vAA}
\end{equation}
where $\Psi _r$ defines the reaction plane.

The azimuthal asymmetry of three-jet events was also observed in $Sp\bar pS$
collider experiments \cite{arn}.

In principle, several other parameters may be used to quantify the azimuthal
asymmetry. In particular, in Dubna experiments the azimuthal coplanarity
was described by the coefficient
\begin{equation}
\beta = \frac {\sum ^n_{i>j} \cos 2 ( \phi _i-\phi _j)}{\sqrt {n(n-1)}},
\label{beta}
\end{equation}
where $\phi _i-\phi _j$ denotes the difference between azimuthal angles of
particles $i$ and $j$.

To describe the alignment phenomenon in cosmic ray data the coplanarity
coefficient was introduced \cite{sla}:
\begin{equation}
\lambda _n = \frac {\sum ^n_{i\neq j\neq k} \cos 2\phi ^k_{ij}}{n(n-1)(n-2)},
\label{lam}
\end{equation}
where $\phi ^k_{ij}$ is the angle between the straight lines connecting the
$i$th and $j$th "particles" (cores)  with the $k$th one.

In emulsion experiments, particle distributions in an individual event were
often shown. Nowadays, experiments at the RHIC \cite{star, phenix, phobos} 
and the LHC \cite{cms} provide results in a form of the
two-dimensional ($\Delta \eta, \; \Delta \phi $) correlation plots and reveal
the ridge phenomenon at large multiplicities. Since the individual events
have no definite orientation in the azimuthal plane, in inclusive one-particle
distributions there is no preferred direction. It appears when the trigger
azimuthal angle is fixed. Thus the elongated individual events become
oriented and the ridge appears. Nevertheless, it is of the same nature as
elliptic flow and alignment and all the above criteria may be applied to
quantify it.

The problem of origin of all these effects is widely debated now, see, e.g., \cite{shu}-\cite{dremin}. 
We would like to notice that it may be ascribed to the general properties of the quantum
field processes. The matrix elements of high energy processes are constructed
in such a way that incoming partons become more and more coplanar with the
outgoing partons \cite{ber, hal}. The key role in them is played by the propagators of 
exchanged partons. The matrix elements are larger where their denominators are smaller.
Considering the parton exchange in the (multi)peripheral (or multi-Regge) kinematics
 at very high energies,  where the exchanged partons
form a string (ladder) with low transferred momenta, the dominant contribution 
to  the processes $ p_A+p_B\rightarrow k_1+k_2+k_3+...$ with three, four, ... produced partons
comes from the  factors $1/t_1t_2, \; 1/t_1t_2t_3,...$ \cite{dre}-\cite{bfkl}. 

For three produced partons this factor is merely
\begin{equation}
\frac {1}{(1-\cos \theta _1)(1+\cos \theta _2)},
\end{equation}
where $\theta _1$ and $\theta _2$ are polar angles to the collision axis.
This factor is large if all three partons tend to be coplanar with the
collision axis.

For four produced partons the factors adjacent to the ends of the graph ($1/t_1$ and $1/t_3$)
are similar to the above expressions, while
the factor $1/t_2$ in between them is to be considered separately. 
Then the following expression appears in the denominator
\begin{eqnarray}
p_Ak_1(1-\cos \theta _1) &+& p_Ak_2(1-\cos \theta _2) \\ 
& - & k_1k_2[1-
\cos (\theta _1-\theta _2)+\sin \theta _1\sin \theta _2(1-\cos (\phi _1-\phi _2))]. \nonumber 
\end{eqnarray}
It is clearly seen that the same tendency to coplanarity is observed with small
differences in polar and azimuthal angles between all the momentum vectors.
Apparently, this asymptotic consideration with $t$-singularities should include
phase space restrictions at finite energies and low-$p_T$ with account of finite parton (jet) masses.

In other words, topology of an inelastic event is basically formed by the plane, 
defined by two final state partons, while the other produced partons are either 
being collinear to them or soft.

Few comments are in order.  The suggested picture assumes that 
not all the produced partons are resolved as (mini-)jets.
Moreover,  in semi-hard regime due to intrinsic momenta of partons inside of colliding hadrons,  
outgoing partons may form only one-side jet-like structure balanced by a back-to-back ridge, 
i. e., unlike hard scattering back-to-back picture the multi-parton cascades
might have more "soft" topology. Also, the considered multi-parton final states can be related 
with high hadron multiplicities and it can cause the near-side bumps in the ridge events 
after multiple parton rescatterings.

To conclude, a quantitative study of coplanarity induced by partonic matrix elements
must be performed before different models are proposed to explain the
observed phenomena. 

We are grateful to V.B. Gavrilov, I.A. Golutvin and  A.A. Vorobyov for useful discussions.
This work was supported in parts by RFBR grants 09-02-0074, 
08-02-91000-CERN, 08-02-01184; RF president grant NS-3383.2010.2 and by the RAS-CERN program.

\end{document}